\documentclass[11pt,preprint]{aastex}
\shorttitle{Novel Technique for Wide-Field Polarimetry}
\shortauthors{McConnell, Carretti and Subrahmanyan}

\newcommand{\lsim}{\raisebox{-0.3ex}{\mbox{$\stackrel{<}{_\sim} \,$}}}
\newcommand{\mbf}[1]{\mbox{\boldmath $#1$}}
\newcommand{\pr}{^{\prime}}

\newcommand{\matb}[2]{
 \left(
   \begin{array}{cc}
     #1 \\
     #2 
   \end{array}
 \right) }
\newcommand{\mata}[4]{
 \left(
   \begin{array}{cc}
     #1 & #2 \\
     #3 & #4
   \end{array}
 \right) }
\newcommand{\matbb}[4]{
 \left(
   \begin{array}{cc}
     #1 \\
     #2 \\
     #3 \\
     #4
   \end{array}
 \right) }
\newcommand{\matbba}[4]{
 \left<\left(
   \begin{array}{cc}
     #1 \\
     #2 \\
     #3 \\
     #4
   \end{array}
 \right)\right> }

\received{2005 May 25}
\begin{document}
%

\title{\large\bf\flushleft A NOVEL TECHNIQUE FOR WIDE-FIELD POLARIMETRY WITH
A RADIOTELESCOPE ARRAY}
\author{D. McConnell}\affil{Australia Telescope National Facility, CSIRO, Epping, NSW, 1710, Australia}
\author{E. Carretti}\affil{INAF-IASF Bologna, Via Gobetti 101, Bologna, I-40129, Italy}
\and \author{R. Subrahmanyan}\affil{Australia Telescope National Facility, CSIRO, Narrabri, NSW, 2390, Australia}
%
%
%

\begin{abstract} We report the use of the Australia Telescope Compact
  Array (ATCA) to conduct polarimetric observations of the sky at
  5~GHz.  The ATCA is normally operated as an interferometer array,
  but these observations were conducted in a split array mode in which
  the antenna elements were used as single-dishes with their beams
  staggered to simultaneously cover a wide area of sky with a
  resolution of 10$\prime$. The linearly polarized sky radiation was
  fully characterized from measurements, made over a range of
  parallactic angles, of the cross correlated signals from the
  orthogonal linear feeds.  We describe the technique and present a
  polarimetric image of the Vela supernova remnant made as a test of
  the method. The development of the techniques was motivated by the
  need for wide-field imaging of the foreground contamination of the
  polarized component of the cosmic microwave background signal.
\end{abstract}


\section{Introduction}

Measurements of the polarized component of the cosmic microwave
background (CMBP) are keenly sought to provide an independent
measurement of cosmological parameters and verify the precision of
certain assumptions that form the basis of cosmological
models. Measurements of the CMBP can also give greater insights into both
the re-ionization and inflation processes than can be gained from
temperature data alone \citep[e.g.][]{zss97,kin99}.  These
measurements are best made at frequencies above 30~GHz and are
technically difficult. At any frequency the measurements are expected
to suffer from foreground contaminants such as Galactic synchrotron
emission and dust \citep[e.g.][]{teh00}, with the synchrotron emission
expected to dominate up to 100~GHz. Observations of Galactic
synchrotron emission at lower frequencies will provide a measure of its
potential to contaminate the CMBP signal.  There have been no
polarization surveys of large areas of sky at frequencies above 2.7
GHz \citep{drr99} and so estimates of contamination at the CMBP
observing frequencies currently rely on extrapolations from that
frequency and from theoretical considerations \citep{c02a}.
Measurements of sky polarization at shorter centimetre wavelengths
would provide useful constraints on that extrapolation.  Surveys at
higher frequencies have been difficult because of the limited fields of
view. In general, the need for wide-field imaging has driven the
development of interferometers with smaller array element sizes and
single-dish telescopes with receiver arrays at the focal plane.  We
describe here a novel method of using existing interferometer arrays
for wide field imaging.  We have used the Australia Telescope Compact
Array (ATCA), configured as a set of single-dishes, to make polarimetric
observations at 5~GHz. Single-dish radio-polarimetry has not
previously been conducted with the ATCA.

The angular power spectrum of the CMB polarization is predicted
to peak around spherical harmonic multipoles $\ell = 1000$,
corresponding to a  $\theta \sim 180^\circ/\ell\sim 11\pr$ angular
scale \citep{zss97}. At 5~GHz (wavelength of 6~cm),  the 22-m ATCA antennas have
 an angular resolution of  FWHM~$\sim 9\pr$.  Therefore,
the 22-m ATCA antennas, operating in single-dish total power mode at
5~GHz, are well suited for  making a survey of the sky polarization at
the multipoles of interest to the planned CMB experiments.

The sky regions that have the lowest dust emission are the Galactic pole
areas.  However, the minimum Galactic synchrotron emission is observed 
at lower latitudes ($|b| = 40^\circ$--50$^\circ$). In particular, 
a region of interest in the southern sky is at RA~$= 5^h \pm 1^h$ 
Dec~$= -45^\circ \pm 10^\circ$ \citep{c02b} which 
culminates at high elevations at southern mid-latitudes
and so is observable with the ATCA antennas  with  low ground spillover.

We describe the methods used at the ATCA for wide-field
radio-polarimetry imaging.  The Array's standard interferometer mode
was used to characterize the instrumental errors and calibrate the
antenna responses, whereas the wide field imaging was done by using
the array in single-dish mode with the pointing of the array elements
offset to instantaneously cover a wide sky area. The method has been
demonstrated using the ATCA to image the Vela supernova remnant.

\section{Method}

The ATCA is a six-element aperture
synthesis radiotelescope that operates at 20, 13, 6, 3 and
1.2~cm and 3.5~mm \citep{fbw92}.  Each element is
a 22-m antenna with (shaped) Cassegrain optics.  Each has three
receiver packages that operate over pairs of wavelength bands.
Receiver tuning and signal sampling are synchronized across the array
by a signal (the local oscillator) distributed from the array
centre.  A range of signal bandwidths can be selected, and each
antenna is equipped to provide two observing bands within the tuning
range of the receiver package in use.  In this work we have used the
6/3~cm package tuned to two 128~MHz bands centred at 4800 and
4928~MHz.  In each band, the signal for each linear polarization is
sampled (2-bit) at the Nyquist rate, and transmitted on optical fibres
to the digital correlator.

In normal operation the correlator computes the cross-correlation of signal
pairs from the array,  the two polarizations from each of the six
elements yielding 60 cross correlations.  These are calibrated and integrated
to form the complex visibilities used in aperture synthesis image formation.
In addition, autocorrelations and polar cross-correlations are computed for
each antenna, and in normal operation are used for calibration.  We
made use of these correlation products for the single-dish
observations described here. 

\subsection{Single-dish polarimetry}
\label{sdp}

The ATCA antennas are on altitude-azimuth mounts and have two orthogonal
linearly polarized feeds $X$ and $Y$ inclined 45$^{\circ}$ to the vertical.
The radiation field incident on the telescope feed has contributions $\mbf{e}$
from the sky, and sources in the immediate environment such as the atmosphere,
the antenna and the ground.  The stray radiation enters the feed both directly
and via a number of reflections off the telescope optics and support
structures.  Although initially unpolarized, response to this radiation in the
$X$ and $Y$ channels may be correlated, resulting in a spurious polarized
signal.  Let $\mbf{s}$ denote the stray radiation and consider it to be
partially polarized.  The telescope optics, even for radiation entering along
the intended path, is imperfect so there is some leakage of each polarized wave
into the other channel.  In this section we follow the treatment of radio
polarimetry by \citet{hbs96}, hereafter HBS96.  We describe
the leakage with the terms $d_x$ and $d_y$ and write a leakage matrix $\mbf{D}$ and
input vectors $\mbf{e}$ and $\mbf{s}$ as
\[ \mbf{D} = \mata{1}{d_x}{-d_y}{1} , \mbf{e} = \matb{e_{x}}{e_{y}} , \mbf{s} =
\matb{s_{x}}{s_{y}} \]
The signal entering the receiver is then
\[ \mbf{D}(\mbf{e}+\mbf{s}) = \mata{1}{d_x}{-d_y}{1}\matb{e_{x}+s_x}{e_{y}+s_y}
\]
The receiver adds noise ($n_x$, $n_y$) to each channel and amplifies by gain
factors ($g_x$, $g_y$).  Writing
\[ \mbf{G} = \mata{g_x}{0}{0}{g_y} , \mbf{n} = \matb{n_{x}}{n_{y}} \]
the input vector $\mbf{v}$ to the correlator
is
\begin{eqnarray}
 \mbf{v} & = & \mbf{G}[\mbf{D}(\mbf{e}+\mbf{s})+\mbf{n}] \\
         & = & \mata{g_x}{0}{0}{g_y} \nonumber
      \matb{e_{x}+s_x+n_x+d_x(e_y+s_y)}{e_{y}+s_y+n_y-d_y(e_x+s_x)}
\end{eqnarray}
The correlator produces the coherency vector which, in the case of single-dish
observations, is the time-averaged outer product (see HBS96) of the input
signal with its complex conjugate
\begin{equation}
 \mbf{V} = \matbb{XX}{XY}{YX}{YY} = \matbba{v_x v_x^{\ast}}{v_x
v_y^{\ast}}{v_y v_x^{\ast}}{v_y v_y^{\ast}}
\end{equation}
Then (adopting a convention where $A_{pq} = a_p a_q^{\ast}$)
\begin{eqnarray*}
XX & = & g_x g_x^{\ast}[e_x e_x^{\ast} + s_x s_x^{\ast} + n_x n_x^{\ast}
           + d_x(e_y e_x^{\ast} + s_y s_x^{\ast}) \\
   & & \mbox{} + d_x^{\ast} (e_x e_y^{\ast} + s_x s_y^{\ast}) \\
   & & \mbox{} + d_x d_x^{\ast}(e_y e_y^{\ast} + s_y s_y^{\ast})] \\
   & = & G_{xx}[E_{xx} + S_{xx} + N_{xx}
           + d_x(E_{yx} + S_{yx}) \\
   & & \mbox{} + d_x^{\ast} (E_{xy} + S_{xy}) \\
   & & \mbox{} + d_x d_x^{\ast}(E_{yy} + S_{yy})]
\end{eqnarray*}
where the $<>$ symbols are omitted but understood and terms such as $e_x
s_x^{\ast}$ and $s_x n_x^{\ast}$ are dropped because their factors are
uncorrelated and the time-averaged products vanish.  Similarly
\begin{eqnarray*}
XY & = & G_{xy}[E_{xy} + S_{xy}
          + d_x(E_{yy} + S_{yy}) \\
   & & \mbox{} - d_y^{\ast} (E_{xx} + S_{xx}) \\
   & & \mbox{} + d_x d_y^{\ast}(E_{yx} + S_{yx})]
\end{eqnarray*}
Note that the receiver noise components $n_x$ and $n_y$ are uncorrelated and so
the term $<N_{xy}>$ vanishes and does not appear in the expression for
$XY$.

For the ATCA antennas, the leakage terms are stable and easily
measured with the Array configured for interferometry.  Moreover, they are
small ($d \lsim 0.03$) and so second order terms ($d_x d_x^{\ast}$, $d_x
d_y^{\ast}$, etc.) can safely be neglected for this experiment.  The
components of  $\mbf{V}$ are then
\begin{eqnarray}
XX & = & G_{xx}[E_{xx} + S_{xx} + N_{xx}
           + d_x(E_{yx} + S_{yx}) 
	   + d_x^{\ast} (E_{xy} + S_{xy})] \\
YY & = & G_{yy}[E_{yy} + S_{yy} + N_{yy}
           + d_y(E_{xy} + S_{xy}) 
	   + d_y^{\ast} (E_{yx} + S_{yx})] \\
XY & = & G_{xy}[E_{xy} + S_{xy}
          + d_x(E_{yy} + S_{yy})
          - d_y^{\ast} (E_{xx} + S_{xx})] \\
YX & = & XY^{\ast}
\end{eqnarray}

The quantities of interest are the Stokes characterisation of the
astronomical signal and can be related to the terms in ($e_x$,$e_y$) in the
expressions above as (see HBS96)
\begin{eqnarray}
E_{xx} & = & \frac{I+Q\pr}{2} \\
E_{xy} & = & \frac{U\pr+iV}{2} \\
E_{yx} & = & \frac{U\pr-iV}{2} \\
E_{yy} & = & \frac{I-Q\pr}{2} 
\end{eqnarray}
We seek to measure the linearly polarized sky emission which is characterized
by $Q$ and $U$. In Equations 7--10 we use the primed symbols $Q\pr$ and $U\pr$ to
emphasize that these are the linear Stokes parameters in the frame of the
antenna feeds, not the required $Q$ and $U$ defined relative to the cardinal
direction on the sky which must be determined by rotation through the
parallactic angle $\psi$ later in the analysis.  Expressed as equivalent
temperatures, the typical size of the terms in Equations 3--6 for the
parameters of our observations are
\begin{eqnarray*}
E_{xx} & \simeq & E_{yy}  \; \simeq \; 5\mbox{K} \\
S_{xx} & \simeq & S_{yy}  \; \simeq \; 5\mbox{K} \\
N_{xx} & \simeq & N_{yy}  \; \simeq \; 25\mbox{K} \\
E_{xy} & \simeq & {E_{xx} - E_{yy}\over 2} \; \simeq \; 0.001\mbox{K} 
\end{eqnarray*}

It can be seen that $U\pr$ can be determined from $XY$, the component
containing terms in $E_{xy}$, whereas the determination of $Q\pr$
involves the difference $XX - YY$. Thus any fluctuations or errors in
the determination of the receiver gain will, when multiplied by the
large terms in Equations 3 and 4, produce large uncertainties in the
value of $Q\pr$.  For this reason single-dish radio polarimetry is
best performed with feeds receptive to opposite hands of circular
polarization so that all the uncertainties arising from differencing
the parallel handed components of the coherency vector ($RR$, $LL$)
flow into the circular Stokes $V$ component which is often  of less
astrophysical interest than the linear polarization.  Radio
interferometry is protected from this because both the stray radiation
$\mbf{s}$ and the receiver noise $\mbf{n}$ entering the two antennas
of each interferometer are uncorrelated.  Thus the ATCA can
successfully use linear feeds for polarimetric interferometry.  For
this experiment, the uncertainties in $Q\pr$ are intolerable and so we
must determine the full linear polarization state from measurements of
$U\pr$ made at several parallactic angles.  This process is described
in section \ref{duq}.

Further practical difficulties arise from the stray radiation which is
partially polarized and is a strong function of antenna elevation, and also has
some azimuthal dependency.  These dependencies have proven impossible to model
and so all our observations have been conducted as ``drift scans'', in which
the antennas are held at a fixed azimuth and elevation, and the measurements
are recorded as the sky drifts past at the sidereal rate.

The analysis above does not include the possibility of further
additive components of stray signal in the polarized outputs. Examples
are the components arising from coupling of receiver noise between the
two polarizations in the orthomode transducerr, and from common mode
noise from the use of common Local Oscillators in the frequency
conversion stages of the receiver.  These are expected to be constant,
independent of antenna pointing, but are difficult to model.
Consequently absolute measurements of the polarized emission are not
attempted.  All observations are differential, with images of the
polarized emission being presented after the subtraction of a baseline
from each scan.

\subsection{ATCA amplitude calibration}
\label{aac}

At the digitisation stage prior to correlation, the receiver output
signals are normalized.  At the ATCA, the correlator measures
correlation coefficients (which take values in the range [-1,1])
and to recover the absolute scaling  of the coherence measurements a
calibration system continously measures and records the system
temperature $T_{sys}$ of each polarization channel against a noise signal of
known temperature $T_{cal}$ which is injected at 45$^{\circ}$ to the
$X$,$Y$ feeds.  Typically $T_{cal} \simeq 2\mbox{K}$ referred to the
face of the feed horn.  The calibration noise signal is switched and a
synchronous demodulator measures the signal power $p$ and $q$ during
the off and on phases respectively. Let $\sigma$ be  the measurement
error on the normalized correlator outputs

\begin{equation}
 \sigma = \frac{\Delta V\pr_i}{V\pr_i} \simeq \frac{1}{\sqrt{\tau B}}
\end{equation}
where $\tau$ and $B$ are the integration time and signal bandwidth,
respectively.  A factor $G$ can then be used to scale the normalised
correlator outputs
\begin{equation}
\mbf{V} = G \mbf{V\pr} = \frac{T_{cal}}{q-p} \mbf{V\pr}
\end{equation}
The measurement error $\mbf{V}$ has contributions from the error
$\Delta G$ in the determination of $G$ and from the noise $\sigma$ on
the normalized correlator outputs
\begin{equation} 
\frac{\Delta G}{G}  =  \left[\frac{\Delta(q-p)}{q-p} \right] \\
\end{equation}
The quantities $p$ and $q$ are independent, being formed by
integrating the
signal over distinct time intervals $\tau/2$.  Let $f =
T_{cal}/T_{sys}$. Then $q = p(1+f)$ and
\begin{eqnarray}
\Delta p      & = & \sqrt{2}\sigma p \nonumber \\
\Delta q      & = & \sqrt{2}\sigma q =  \sqrt{2}\sigma p(1+f)
\nonumber \\
\Delta ( q-p) & = & \sqrt{(\Delta q)^2 + (\Delta p)^2}  \nonumber \\
              & =  & 2\sigma p\sqrt{1+f+f^2/2}  \nonumber \\
\frac{\Delta(q-p)}{q-p}  & =  & \frac{2 \sigma \sqrt{1+f+f^2/2}}{f}
\end{eqnarray}
Since at the ATCA $f \simeq 0.05$, the errors in the $V_i$ are
dominated by the error in the determination of $G$
\begin{eqnarray}
\frac{\Delta V_i}{V_i} & \simeq & \frac{2 \sigma \sqrt{1+f}}{f} \nonumber \\
 & \simeq & 40\sigma \; \mbox{\hspace{5em}when $f = 0.05$}
\end{eqnarray}
Thus, the application of the online $T_{sys}$ calibration increases
the noise in $\mbf{V}$ by a factor of about $2/f$ and for this
experiment severely limits the sensitivity of the measurements.  In
addition, since the calibration signal is injected at $45^{\circ}$ to
the $X$,$Y$ feeds, it appears as an additional strong polarized
component in Equation 5 which, when multiplied by uncertainties and
fluctuations in $G_{xy}$ could overwhelm the polarized signal from the
sky.  Our observations were made with the online application of
$T_{cal}$ scaling disabled. The scaling factor $G$ was determined
independently from separate integrations that bracketted the observations
of the sky.

Note that the difference in the instrumental path lengths for the $X$
and $Y$ polarizations is expected to be non-zero, and antenna specific but
only slowly changing. Failing to account for it rotates the signal between
Stokes $U$ and $V$. We measured this path-length (phase) difference for
each antenna by measuring the phase of the calibration signal that is
injected at $45^{\circ}$ to the $X$,$Y$ feeds. The phase correction
was then applied to the measured $XY$ values.

\subsection{Observational procedure}

To measure the polarized emission from a $\Delta \alpha \times \Delta \delta$
rectangular patch of sky centred at $(\alpha,\delta)$ we performed a series of
drift scans of duration $\Delta t = \Delta \alpha/ \cos \delta$, with the six
antennas pointed at the same hour angle and each offset in declination from the
next by half the width of the primary beam $\theta$.  In
successive scans the declination was incremented by $3\theta$ and
the hour angle changed (by approximately $\Delta t$) to scan the same range of
right ascensions. The entire patch was covered after

$n_s = \Delta \delta/3\theta$

scans, and then repeated for the
required total integration time.  In this way each point in the surveyed area
was measured at intervals of approximately
\[
n_s \Delta t = \frac{\Delta\alpha \Delta\delta}{3\theta \cos
\delta}
\]

\subsection{Determination of $Q$ and $U$}
\label{duq}

As described in section \ref{sdp}, the ATCA is able to provide only one
correlated output of the linear polarization Stokes parameters, $U\pr$. Thus
$Q$~and~$U$ (components defined in sky coordinates) must be constructed starting
from  this single quantity $U\pr$ by performing at least two scans at
parallactic angles $\psi$ ideally differing by $\pm 45^\circ$.

The general case consists of $N$ observations of the same pixel
performed at several parallactic angles.  The two Stokes parameters
are estimated through a least square approach. Let ${\tilde U\pr}_i$
be the $i^{th}$~observation of $U\pr$. Its expression in terms of the
actual $Q$ and $U$ values of the sky in the standard reference frame
is
\begin{equation}
  {\tilde U}\pr_i = -Q\,\sin(2\,\psi_i) + U\,\cos(2\,\psi_i).
\end{equation}
where $\psi_i$ is the parallactic angle during the observation.  (In
Equation 17, $\psi_i = p_i - \phi$ where $p_i$ is the parallactic
angle and $\phi$ is the fixed angle between the feeds and the
vertical-horizontal directions. At the ATCA $\phi = 45^{\circ}$.)
Minimizing the sum of the square differences
\begin{equation}
  S^2 = \sum_i \left( {\tilde U}\pr_i - U\pr_i\right)^2
\end{equation}
with respect to both $Q$ and $U$, we obtain as best estimates
\begin{eqnarray}
    && \nonumber\\ 
  Q &=& {  \sum_i {\tilde U}\pr_i\,\sin(2\,\psi_i) \sum_i\cos^2(2\,\psi_i) - 
           \sum_i {\tilde U}\pr_i\,\cos(2\,\psi_i) \sum_i\sin(2\,\psi_i)\cos(2\,\psi_i) 
         \over
           \left(\sum_i \sin(2\,\psi_i)\cos(2\,\psi_i)\right)^2 -
           \sum_i\sin^2(2\,\psi_i)\sum_i\cos^2(2\,\psi_i)} 
           \label{Qeq}\\
    && \nonumber\\ 
    && \nonumber\\ 
  U &=& {  \sum_i {\tilde U}\pr_i\,\sin(2\,\psi_i) \sum_i\sin(2\,\psi_i)\cos(2\,\psi_i) - 
           \sum_i {\tilde U}\pr_i\,\cos(2\,\psi_i) \sum_i\sin^2(2\,\psi_i) 
         \over
           \left(\sum_i \sin(2\,\psi_i)\cos(2\,\psi_i)\right)^2 -
           \sum_i\sin^2(2\,\psi_i)\sum_i\cos^2(2\,\psi_i)}
           \label{Ueq}\\
    && \nonumber 
\end{eqnarray}
whose errors, assuming all the samplings have the same sensitivity $\sigma_i$, 
are
\begin{eqnarray}
    && \nonumber\\
  {\sigma^2_Q / \sigma_i^2}
         &=& {  \sum_i \sin^2(2\,\psi_i)
                    \left[\sum_i\cos^2(2\,\psi_i)\right]^2 -
                    \sum_i \cos^2(2\,\psi_i)
                    \left[\sum_i\sin(2\,\psi_i)\cos(2\,\psi_i)\right]^2
         \over
                    \left[
                         \left(\sum_i \sin(2\,\psi_i)
                                      \cos(2\,\psi_i)
                         \right)^2 -
                         \sum_i\sin^2(2\,\psi_i)
                         \sum_i\cos^2(2\,\psi_i)
                    \right]^2}
                    \label{sQeq}\\
    && \nonumber\\
    && \nonumber\\
  {\sigma^2_U / \sigma_i^2}
             &=& { -\sum_i \sin^2(2\,\psi_i)
                    \left[\sum_i\sin(2\,\psi_i)\cos(2\,\psi_i)\right]^2 +
                    \sum_i \cos^2(2\,\psi_i)
                    \left[\sum_i\sin^2(2\,\psi_i)\right]^2
         \over
                    \left[
                         \left(\sum_i \sin(2\,\psi_i)
                                      \cos(2\,\psi_i)
                         \right)^2 -
                         \sum_i\sin^2(2\,\psi_i)
                         \sum_i\cos^2(2\,\psi_i)
                    \right]^2}
                    \label{sUeq}\\
    && \nonumber
\end{eqnarray}
giving the fractional errors with respect to the sensitivity of a
single observation.

\begin{figure*}
\epsscale{.95}
\plotone {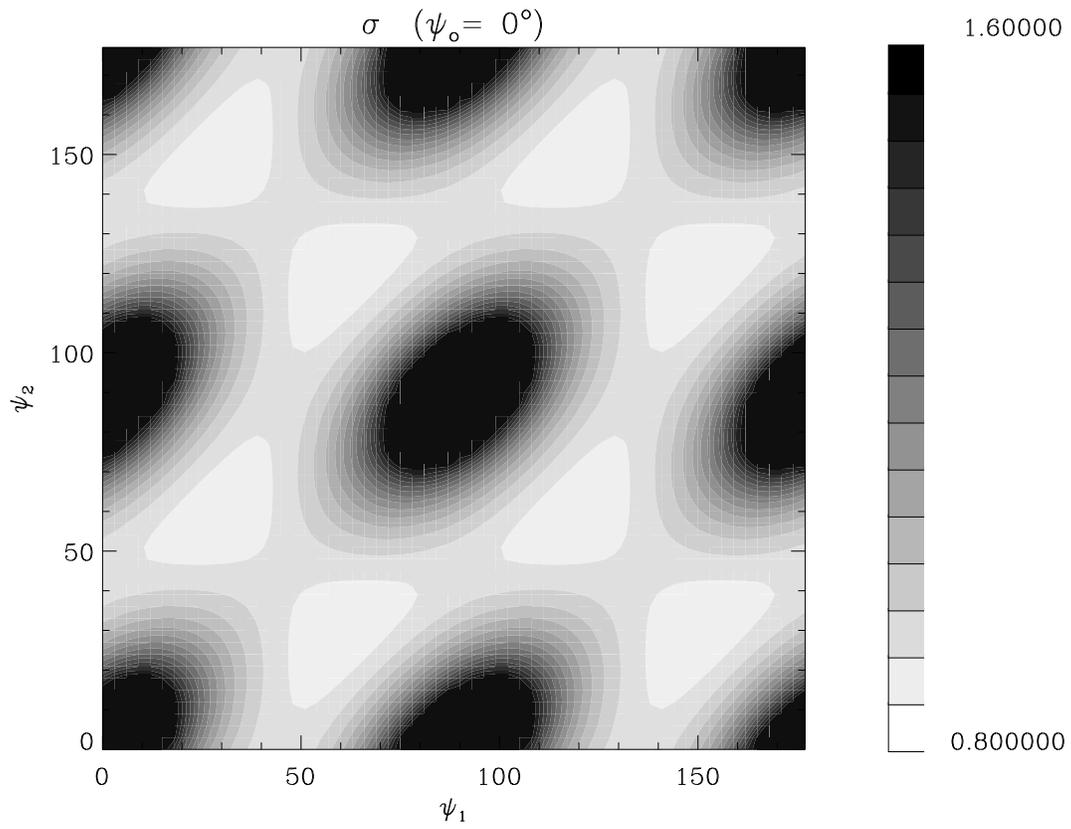}
\caption{Behaviour of the errors
           $\sigma / \sigma_i$ for the case with $N=3$: The value of $\psi_0$ is
           kept constant, while $\psi_1$ and $\psi_2$ vary over the
           range 0$^\circ$ to 180$^\circ$.}
  \label{sigma3PAFig}
\end{figure*}
These errors depend on the set of parallactic angles at which the
measurements occur. The efficiency of the reconstruction thus depends
on the scanning strategy adopted, which must be optimized to minimize
the error on both $Q$ and $U$.  Considering that the method provides
two Stokes parameters from measurements of only one, the ideal
sensitivity for $N$ observations is
\begin{equation}
   {\sigma^{\rm th}_{Q,U} \over \sigma_i} = \sqrt{2\over N}.
\end{equation}
Its comparison with Equations~(\ref{sQeq}) and~(\ref{sUeq})
allows a quantitative estimate of how well $Q$ and $U$
are measured.
In addition, to evaluate the overall effect on the ($Q$, $U$) pair,
we introduce the mean error 
\begin{equation}
   {\sigma \over  \sigma_i} = \sqrt{{1\over 2}{\sigma_Q^2 + \sigma_U^2 
                                               \over \sigma_i^2}}.
\end{equation}

For the simplest case with N=2 observations, the best result is obtained when 
the difference  ${\rm \psi}_1 - {\rm \psi}_0 = \pm\,45^\circ$, as expected.

The case N=3 is less trivial. Keeping in mind the ideal
value ($\sigma / \sigma_i = 0.82$ for $N=3$), the case is clear
looking at Figure~\ref{sigma3PAFig}, where $\psi_0$ is kept constant,
while the other two angles $\psi_1$ and $\psi_2$ are allowed to vary
in the 0$^\circ$--180$^\circ$ range.  The best sensitivity occurs when
the three angles are evenly separated in a $90^\circ$ interval,
rather than differing by $45^\circ$.  The case with $\pm 45^\circ$
separations favours one of the two Stokes parameters,
resulting in a non-optimal combined sensitivity $\sigma$.
In general, the optimum result from $N$ observations is obtained with
the $90^\circ$ range sampled with equally spaced parallactic angles.

\section{Test observations}

To provide some confidence in the method, we observed $3.5 \times 3.0$ degree
region centred on the Vela supernova remnant for about 10~hr. This region
emits strong polarized emission and has been well studied at a range of
wavelengths and resolutions.  \citet{m80,m95} reported polarimetric
observations of the Vela nebula at 2.7GHz (resolution 8.4 arcminutes), 5.0 GHz
(4.4 arcminutes) and 8.4~GHz (3 arcminutes). \citet{dhj97} measured the
polarized emission from the southern Galactic plane at 2.4~GHz (resolution 10.4
arcminutes), including the Vela region.

We observed the area with repeated 20-minute drift scans.  Each scan
was started with all antennas set to $\alpha = 08^{h}27^{m}$ and
declinations spaced by 5 arcminutes.  Thus each
scan sampled a 0.5 degree declination band, and six scans were
required to sample the whole area.  The six scans were repeated for
10~hr, allowing each point in the surveyed area to be measured
four or five times.  The scan data were reduced to form the image
shown in Fig. 2.  As described in section \ref{aac}, the ATCA flux
scale is referred to a switched noise signal, the ``on-line $T_{sys}$
measurement''.  The total intensity image was derived from those
measurements.    The full recovery of $Q$ and $U$ required each point
to be measured at different parallactic angles (see section
\ref{duq}).  In this case the south-west corner of the surveyed area
was observed five times at parallactic angles of $-117$, $-96$, $-85$,
$-70$ and $-48$ degrees.

\begin{figure*}
\epsscale{0.95}
\plotone {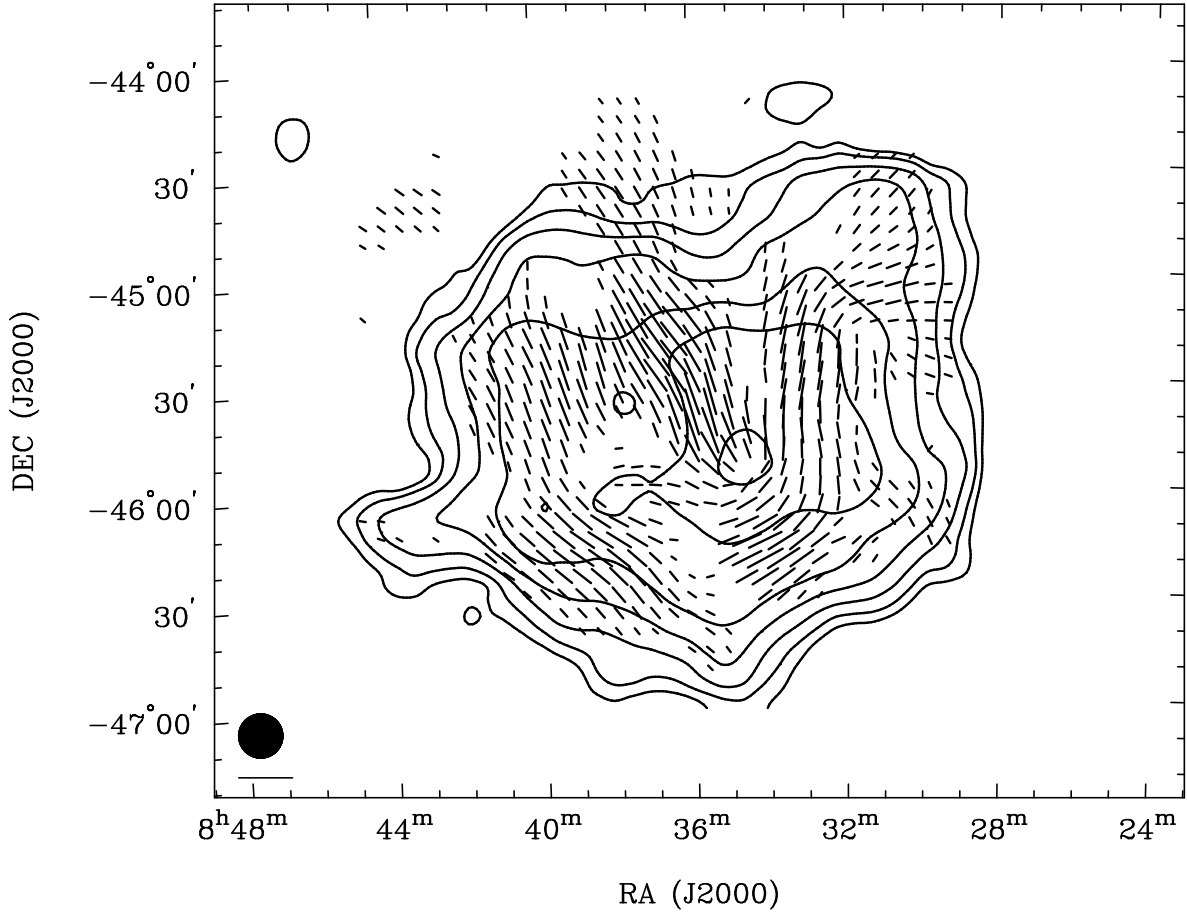}
  \caption{The Vela supernova remnant at 4.9~GHz.  The image has a
  resolution of 12$\prime$ (indicated at lower left).  The 4.9~GHz data
  were combined with the published 2.4~GHz image (Duncan et al., 1997)
  to determine the rotation measure across the nebula, and thus the
  intrinsic polarization position angles.  Total intensity is shown as
  contours (0.06, 0.10, 0.17, 0.27, 0.44, 0.72, and 1.16~K), polarized
  intensity and magnetic field direction are shown as line segment lengths and
  angles, respectively. The scale of the line segment lengths is indicated by the
  segment at lower left which is equivalent to 0.19~K.}
 \label{velaFig}

\end{figure*}

Figure \ref{velaFig} shows the resulting image of the Vela supernova
remnant.  The peak total and polarized intensities are 1.3~K and
0.19~K respectively.  The image has a resolution of 12 arcmin.  Our
image compares well with published data and verifies our techniques
for surveying an area in single-dish mode with series of drift scans
and the complete measurement of Stokes $Q$ and $U$ through sampling a
range of parallactic angles.  In particular there is a good match of
the polarization position angles between our results and those of
\citet{m80,m95}.  This indicates successful removal of the
polarization offsets and background polarization which, as reported by
\citet{m80} is low relative to the polarization of the nebula itself.

The high brightness of the Vela region makes our test image less
useful for assessing the ultimate sensitivity of the method and its
ability to measure the weak CMBP foregrounds. The apparent noise in
the $Q$ and $U$ images of $\sigma_{Q,U} \simeq 3 \mbox{~mK}$ is
dominated by the variations in polarized emission over the field.
However, we have made preliminary observations of a region near
$\alpha = 5^{\rm h}$,~$\delta = -49^{\circ}$ at high Galactic
latitude, which is expected to have low foreground emission.  This is
the region chosen for a number of CMBP measurements (e.g. BaR-SPOrt,
\citealt{cortiglioni03}, BOOMERanG-B2K, \citealt{masi05}).  We made
the high-latitude observations using the same method described above
for the Vela supernova remnant and used a total bandwidth of
$\sim$200~MHz.  The results of the completed 5~GHz ATCA observations
of this region will be the subject of a future report, in which we
will compare them with equivalent measurements at lower
frequencies. Reducing the preliminary measurements to determine the
sensitivity of a single telescope over a 1-s integration we find
$\sigma_Q = 5.8$ and $\sigma_U = 5.3\mbox{~mK s}^{1/2}$, about a factor
of 1.6 greater than expected from an ideal noise analysis.  We can
estimate the sensitivity to $Q$ and $U$ for a survey area of $\Omega$
with angular resolution $\Delta\Omega$ using $n_t$ telescopes as
\begin{equation} \sigma_{Q,U} \simeq \frac{5.5}{\sqrt{n_t T
\Delta\Omega/\Omega}} \mbox{~mK s}^{1/2}
\end{equation} where $T$ is the total integration time for the survey.
For example a 10-hour observation using all six ATCA telescopes should
yield $\sigma_{Q,U} \simeq 0.06 \mbox{ mK}$ for a survey area of 1 deg$^2$
and a resolution of 12$\prime$. This sensitivity
would allow a 3$\sigma$ detection of the expected 5~GHz signal of
$\sim$0.2\,mK (see \citet{c05} for measurements of the polarized
foreground at 2.3~GHz).

\section{Summary}

We have described a novel method, and its demonstration, of using an
existing interferometer array for wide field imaging. The
interferometer mode of observing was used to derive the instrumental
calibration and a single-dish observing mode with pointing offsets
between the array elements was used for the wide field observing. We
used the one correlated output ($U\pr$) to construct both $Q$ and $U$
by observing all points in the imaged area at several parallactic
angles.  We recognise this as a potential technique for all-sky
surveying or monitoring with future radio telescope arrays such as the
Square Kilimetre Array (SKA) if the array elements are chosen to have
limited sky coverage.  This could relax the constraint that the SKA must
be built with wide-field elements and allow more conventional
parabolic dishes to be used as the SKA elements.

\section*{Acknowledgments} 

We are grateful to M.H. Wieringa for assistance with modifications to
the ATCA control software, and to other members of the ATCA staff for
expert support during the development of this observational mode. We
thank Simon Johnston for reading the manuscript and suggesting several
improvements. We also thank the referee for suggestions that have led
to improvements in the manuscript. The ATCA is funded by the
Commonwealth of Australia for operation as a National Facility by
CSIRO.

\end{document}